\documentclass[10pt,conference]{IEEEtran}
\usepackage[utf8]{inputenc}
\usepackage{amsmath,amssymb}
\usepackage{graphicx}
\usepackage{algorithm}
\usepackage{algpseudocode}
\usepackage{booktabs}
\usepackage{multirow}
\usepackage{comment}
\usepackage{xcolor}
\usepackage{hyperref}
\usepackage{tabularx}
\hypersetup{
    colorlinks=true,
    linkcolor=blue,
    filecolor=magenta,      
    urlcolor=cyan,
}
\usepackage{tikz}
\usetikzlibrary{shapes,arrows,positioning,shadows,decorations.pathreplacing,calc}
\usepackage{pgfplots}
\pgfplotsset{compat=1.18}
\usetikzlibrary{intersections} \usepgfplotslibrary{fillbetween}

\usepackage{verbatim}
\newcommand\copyrighttext{%
  \footnotesize \textcopyright 2026 IEEE. Personal use of this material is permitted.
  Permission from IEEE must be obtained for all other uses, in any current or future
  media, including reprinting/republishing this material for advertising or promotional
  purposes, creating new collective works, for resale or redistribution to servers or
  lists, or reuse of any copyrighted component of this work in other works.}
\newcommand\copyrightnotice{%
\begin{tikzpicture}[remember picture,overlay]
\node[anchor=south,yshift=10pt] at (current page.south) 
  {\fbox{\parbox{\dimexpr\textwidth-\fboxsep-\fboxrule\relax}{\copyrighttext}}};
\end{tikzpicture}%
}

\title{The Quantum Learning Pyramid (QLP): A Novel, Holistic, Industry-Ready Curriculum and Pedagogical Methodology for Quantum Computing Education}

\author{\IEEEauthorblockN{Arun Govindankutty}
\IEEEauthorblockA{\textit{Electrical \& Computer Engineering,} \textit{North Dakota State University}
Fargo, ND USA}
}

\begin{document}

\maketitle
\copyrightnotice
\begin{abstract}
Quantum computing education is becoming urgent as industry demand and national initiatives grow rapidly. This paper introduces the Quantum Learning Pyramid (QLP), a unified pedagogical framework for undergraduate and graduate education in quantum information and computing. The QLP follows a four-tier structure that integrates phenomenological understanding, computational thinking, hardware-aware development, and societal context. The curriculum is designed using spiral progression, competency-based pathways, and authentic assessment. Instruction is grounded in active and project-based learning, aligned with ACM/IEEE curriculum guidelines. Core topics include quantum mechanics fundamentals, qubit operations, and key algorithms, while advanced modules address error correction, cryptography, and quantum hardware. Hands-on learning is supported through simulation platforms, cloud-accessible quantum processors, and hybrid laboratory environments. Interdisciplinary case studies and real-system experimentation are embedded throughout. The proposed framework bridges theory and practice and provides a scalable roadmap for developing a quantum-ready workforce and scientifically informed citizens.

\end{abstract}


\section{Introduction}
Quantum information science and computing (QIS\&C) is advancing at an unprecedented pace. Progress in algorithms, hardware platforms, and national-scale initiatives has moved quantum computing from theory to practice~\cite{fernandez2020towards, arute2019quantum, choi2023preparing, zhong2020quantum, qc_arun, AbuGhanem_2025}. As a result, the need for effective and scalable quantum education has become urgent. However, most existing educational models were not designed for the interdisciplinary and computational nature of modern quantum technologies.

Traditional quantum mechanics instruction was developed primarily for physics majors in the twentieth century~\cite{Norsen2017}. While mathematically rigorous, this approach poses significant barriers for today’s learners. Quantum phenomena such as superposition and entanglement are highly counter-intuitive. The heavy reliance on linear algebra and complex analysis further limits accessibility~\cite{nielsen2010quantum, Johnston2019}. In addition, current curricula often treat physics, computer science, and engineering as separate domains, creating gaps between theory, software, and hardware perspectives. These challenges are amplified by the rapid evolution of the field and the wide diversity of student backgrounds, ranging from theoretical science to applied engineering. Thus current challenges can be summarised as:
\begin{enumerate}
    \item Conceptual barriers
    \item Mathematical prerequisites
    \item Interdisciplinary gaps
    \item Rapidly evolving field
    \item Diverse learner needs
\end{enumerate}

Existing educational responses have been largely incremental. Common approaches include adding isolated quantum modules to classical courses or offering specialized graduate programs~\cite{hu2024}. While valuable, these efforts do not adequately address the full spectrum of learner needs or provide a coherent pathway from fundamentals to practice~\cite{garces2025introducingquantumcomputinghighschool}. As a result, many graduates remain unprepared for emerging roles in quantum information science and technology.

At the same time, workforce demand continues to grow. Recent reports indicate that job openings in quantum computing significantly exceed the number of qualified candidates~\cite{BaoTran2026}. Most computing programs still offer limited exposure to quantum concepts. Recognizing this gap, the ACM/IEEE CS2023 guidelines~\cite{Kumar2024} introduced new learning outcomes and a dedicated knowledge unit on quantum architectures. These developments provide an important foundation but also highlight the need for integrated pedagogical frameworks that align theory, practice, and industry relevance.

This paper introduces the \emph{Quantum Learning Pyramid (QLP)}, a comprehensive educational framework designed to address these challenges. The QLP is structured as a four-tier model that integrates phenomenological learning, computational thinking, hardware-aware development, and sociotechnical context. Rather than treating these dimensions independently, the framework connects them through spiral progression and competency-based pathways. This design allows students to revisit core concepts at increasing levels of depth while developing practical skills.

The proposed curriculum emphasizes active and project-based learning. Abstract quantum concepts are reinforced through hands-on experimentation using simulation tools and cloud-accessible quantum processors. Laboratory activities and open-ended projects enable students to explore algorithms, analyze hardware constraints, and evaluate real-world applications. Formal technical communication and interdisciplinary collaboration are embedded throughout, reflecting authentic research and industry practice.

The QLP is modular and adaptable. It supports customization across undergraduate and graduate levels, course durations, and student backgrounds. In the remainder of this paper, we describe the curriculum architecture, teaching methodology, minimum viable implementation, and alignment with ACM/IEEE educational guidelines. Together, these elements provide a scalable roadmap for educating a quantum-ready workforce and fostering scientifically literate citizens.

\section{Theoretical Foundation}
\label{sec:theory}

The Quantum Learning Pyramid (QLP) is grounded in established learning science while being tailored to the unique cognitive demands of quantum information science. The framework integrates constructivist learning theory~\cite{andrew2022}, threshold concepts, and deliberate practice to support deep conceptual understanding and sustained skill development. These foundations guide both curriculum structure and instructional design.

\subsection{Pedagogical Principles}
Quantum learning is characterized by abstract concepts, non-classical reasoning, and high mathematical density~\cite{Norsen2017}. To address these challenges, the QLP emphasizes concept formation before formalism, repeated exposure across contexts, and active engagement through practice and reflection. This design choice draws on discipline-based education research across physics, engineering, and computing: active learning has been shown to raise examination performance and reduce failure rates relative to traditional lecturing across STEM disciplines~\cite{Freeman2014}, conceptual understanding and problem solving in physics are governed by well-documented cognitive mechanisms~\cite{Docktor2014}, and computing education research has independently identified troublesome, threshold concepts in programming and computational thinking that parallel those found in physics~\cite{SandersMcCartney2016}. The QLP's threshold-concept and cognitive-load framing is intended as an application of these established findings to the specific demands of quantum information science and computing. 
\subsubsection{Quantum Threshold Concepts}
Certain ideas in quantum computing act as conceptual gateways. Until these ideas are understood, learners struggle to make progress. We identify the following threshold concepts as central to quantum education:
\begin{center}
$T_{\text{quantum}} = \{ \psi\text{-epistemic shift},\ \text{non-commutativity},\newline 
\text{superposition as basis},\ \text{entanglement as resource},$
\newline
$\text{measurement disturbance},\ \text{quantum--classical interface},\newline \text{no-cloning} \}$
\end{center}

Each threshold concept requires targeted instructional strategies. For example, the $\psi$-epistemic shift, that is moving from classical state descriptions to quantum state representations, needs to be supported using analogy mapping, visual state representations, and side-by-side classical comparisons. These approaches help students restructure their mental models rather than memorize formal rules. Readers are referred to~\cite{Griffiths1995} for detailed technical discussions of the threshold concepts, which are beyond the scope of this work.

\subsubsection{Cognitive Load Management}

Quantum topics impose a high intrinsic cognitive load due to abstract representations and unfamiliar logic. The QLP applies Cognitive Load Theory~\cite{Sweller2010} to manage this complexity:
\[
CL_{\text{total}} = CL_{\text{intrinsic}} + CL_{\text{extraneous}} + CL_{\text{germane}}
\]

Extraneous load is reduced through scaffolded visualizations, stepwise derivations, and tool supported experimentation. Germane load is increased through interleaved practice, reflection activities, and problem variation~\cite{Parveen2025}. 
This balance allows students to focus cognitive effort on building durable conceptual understanding.

\subsection{Learning Progression Model}

The QLP operationalizes learning through a Spiral Quantum Curriculum (SQC). Core concepts are revisited multiple times at increasing levels of abstraction and technical depth. This spiral structure supports students with diverse backgrounds while enabling advanced learners to progress rapidly.

\begin{table}[!h]
\caption{Spiral Progression of Core Quantum Concepts}
\centering
\begin{tabularx}{0.48\textwidth}{
  >{\raggedright\arraybackslash}p{0.07\textwidth}
  >{\raggedright\arraybackslash}X
  >{\raggedright\arraybackslash}X
  >{\raggedright\arraybackslash}X
  >{\raggedright\arraybackslash}X
}
\toprule
\textbf{Level} & \textbf{Superposition} & \textbf{Entanglement} & \textbf{Qubits} & \textbf{Algorithms} \\ \toprule
Foundational & Classical analogies & Correlated systems & Physical intuition & Classical motivation \\ \hline
Operational & State vectors & Bell states & Mathematical models & Simple circuits \\ \hline
Algorithmic & Basis changes & Two-qubit gates & Logical operations & Grover \& Shor \\ \hline
Advanced & Density matrices & Multipartite systems & Error correction & NISQ algorithms \\ \bottomrule
\end{tabularx}
\label{tab:sqc}
\end{table}

This progression (Table~\ref{tab:sqc}) ensures that early exposure emphasizes intuition and motivation, while later stages introduce formalism, optimization, and system-level trade-offs.

\subsection{Active and Experiential Learning}

Active learning is central to the QLP. Project-based learning (PBL) connects abstract theory to real-world applications and interdisciplinary domains. Students engage in open-ended projects that require problem formulation, algorithm selection, implementation, and evaluation. For example, learners may explore the application of the HHL algorithm to medical imaging or optimization problems~\cite{Sambhaje2024}. Such projects make relevance explicit and strengthen conceptual retention.

Laboratory exercises are integrated throughout the curriculum. Weekly labs use quantum software platforms such as Qiskit~\cite{javadiabhari2024quantumcomputingqiskit} or Cirq~\cite{Cirq_Developers_2025} to reinforce lecture material. Early labs focus on circuit construction and measurement statistics, such as preparing Bell states and analyzing correlations. Advanced labs require students to debug circuits, implement complete algorithms, and evaluate performance under noise. This progressive lab structure builds confidence and practical competence.

Formal technical communication is also embedded in the learning process. Students produce short conference-style reports describing their project work and participate in structured peer review. This practice reinforces understanding, improves clarity of reasoning, and mirrors professional research workflows. Many projects are co-mentored across disciplines, such as computer science and electrical engineering, to reflect the collaborative nature of quantum research and development.

Additional active learning techniques, including in-class problem solving, flipped lectures~\cite{Lancaster2016}, and interactive simulations are used to support non-intuitive concepts. Visual tools such as Bloch spheres~\cite{Mosseri2001}, probability distributions, and circuit animations help connect mathematical expressions to observable behaviour. Interactive platforms and tutorials are assigned to encourage exploration beyond the classroom.

Together, these theoretical and instructional elements form a cohesive foundation for the Quantum Learning Pyramid, supporting scalable, inclusive, and practice oriented quantum education.

\section{The Quantum Learning Pyramid Framework}
\label{sec:qlp_frmwrk}

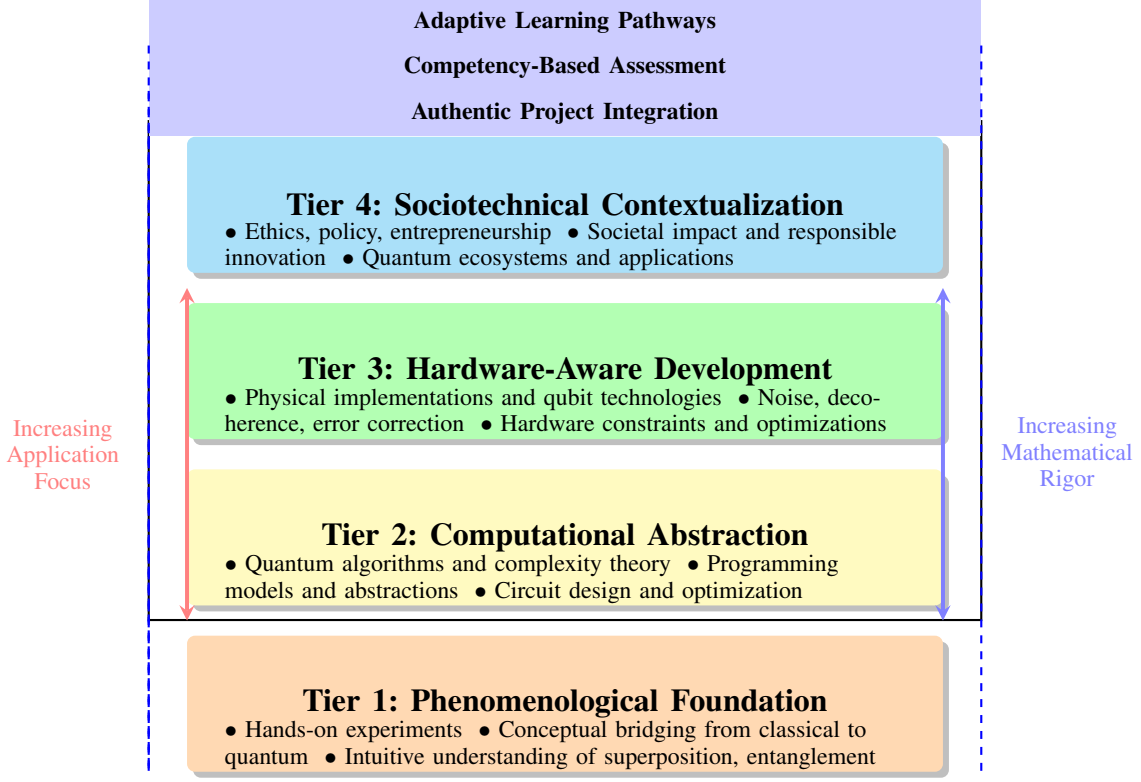
\begin{figure*}[h]
\centering
\begin{tikzpicture}[
    level/.style={rectangle, rounded corners=3pt, minimum width=10cm, minimum height=1.8cm, text centered, drop shadow, font=\large\bfseries},
    tier1/.style={level, fill=cyan!30, text=black},
    tier2/.style={level, fill=green!30, text=black},
    tier3/.style={level, fill=yellow!30, text=black},
    tier4/.style={level, fill=orange!30, text=black},
    cross/.style={rectangle, minimum width=11cm, minimum height=0.6cm, text centered, font=\small},
    arrow/.style={->, >=stealth, thick},
    desc/.style={text width=9cm, align=left, font=\small}
]

\node[tier1] (t1) at (0,0) {Tier 4: Sociotechnical Contextualization};
\node[tier2] (t2) at (0,-2.2) {Tier 3: Hardware-Aware Development};
\node[tier3] (t3) at (0,-4.4) {Tier 2: Computational Abstraction};
\node[tier4] (t4) at (0,-6.6) {Tier 1: Phenomenological Foundation};

\draw[thick] ($(t4.north west) + (-0.5,0.2)$) -- ($(t1.north west) + (-0.5,0.2)$) -- 
              ($(t1.north east) + (0.5,0.2)$) -- ($(t4.north east) + (0.5,0.2)$) -- cycle;

\node[desc, below=-0.8cm of t1] (d1) {
    $\bullet$ Ethics, policy, entrepreneurship \ 
    $\bullet$ Societal impact and responsible innovation \ 
    $\bullet$ Quantum ecosystems and applications
};

\node[desc, below=-0.8cm of t2] (d2) {
    $\bullet$ Physical implementations and qubit technologies \
    $\bullet$ Noise, decoherence, error correction \
    $\bullet$ Hardware constraints and optimizations
};

\node[desc, below=-0.8cm of t3] (d3) {
    $\bullet$ Quantum algorithms and complexity theory \
    $\bullet$ Programming models and abstractions \
    $\bullet$ Circuit design and optimization
};

\node[desc, below=-0.8cm of t4] (d4) {
    $\bullet$ Hands-on experiments \
    $\bullet$ Conceptual bridging from classical to quantum \
    $\bullet$ Intuitive understanding of superposition, entanglement
};

\node[cross, fill=blue!20, above=1.2cm of t1] (c1) {\textbf{Adaptive Learning Pathways}};
\node[cross, fill=blue!20, above=0.6cm of t1] (c2) {\textbf{Competency-Based Assessment}};
\node[cross, fill=blue!20, above=0cm of t1] (c3) {\textbf{Authentic Project Integration}};

\draw[dashed, blue, thick] (c1.south west) -- (c1.south west |- t4.south);
\draw[dashed, blue, thick] (c1.south east) -- (c1.south east |- t4.south);
\draw[dashed, blue, thick] (c2.south west) -- (c2.south west |- t4.south);
\draw[dashed, blue, thick] (c3.south west) -- (c3.south west |- t4.south);

\draw[arrow, <->, blue!50, line width=1.5pt] (5, -1.1) -- (5, -5.5) 
    node[midway, right, text width=3cm, align=center, font=\small] 
    {Increasing \\ Mathematical \\ Rigor};

\draw[arrow, <->, red!50, line width=1.5pt] (-5, -1.1) -- (-5, -5.5) 
    node[midway, left, text width=3cm, align=center, font=\small] 
    {Increasing \\ Application \\ Focus};

\node[above=0.5cm of c1, font=\LARGE\bfseries] {Quantum Learning Pyramid (QLP)};

\end{tikzpicture}
\caption{The Quantum Learning Pyramid (QLP) framework showing four tiers and cross-cutting elements.}
\label{fig:qlp}
\end{figure*}

Building on the pedagogical principles and learning progression described in Section~\ref{sec:theory}, this section presents the Quantum Learning Pyramid (QLP) as an integrated curricular and instructional framework. The QLP organizes quantum education into four vertically aligned tiers, supported by cross-cutting mechanisms for assessment, personalization, and authentic practice. Figure~\ref{fig:qlp} illustrates the overall structure.

The pyramid is designed to support learners with diverse backgrounds while enabling systematic progression from intuition to abstraction, implementation, and societal understanding. 
Movement upward through the pyramid corresponds to increasing mathematical rigour and system level reasoning, while horizontal engagement emphasizes real world application and relevance.

\subsection{Tier 1: Phenomenological Foundation}

Tier~1 establishes intuitive understanding by grounding instruction in observable quantum phenomena. Rather than beginning with formal postulates or linear algebra, learners first encounter quantum behaviour through experiments, simulations, and qualitative reasoning. This approach reduces initial cognitive barriers and supports the threshold concept transition from classical to quantum thinking.

Students engage in hands-on or remotely accessible activities such as single-photon interference~\cite{rueckner1996lecture}, Bell inequality~\cite{Peres1999} violations, and quantum random number generation~\cite{Ma2016}. 
These experiences make abstract ideas tangible and motivate subsequent formalization. 
Conceptual bridging is emphasized throughout, linking classical notions to their quantum counterparts, such as probabilities to probability amplitudes, classical bits to qubits via the Bloch sphere, and Boolean logic to quantum gates.

The goal of this tier is not mathematical mastery, but conceptual readiness. 
By the end of Tier~1, students can reason qualitatively about superposition, entanglement, and measurement, and can articulate how quantum behaviour departs from classical intuition.

\subsection{Tier 2: Computational Abstraction}

Tier~2 builds on phenomenological insight to develop quantum computational thinking. 
Here, learners engage with quantum algorithms, circuit models, and programming abstractions that formalize previously observed behaviour. 
The emphasis is on algorithmic reasoning rather than hardware detail.

Instruction follows a structured programming progression, beginning with visual circuit design tools~\cite{HAGHPARAST2025102383} and advancing toward text based quantum programming environments. 
Students implement simple circuits, explore basis changes, and analyse algorithmic primitives such as oracle construction~\cite{tambde2021programminglanguagequantumoracle} and amplitude amplification~\cite{Brassard_2002}. 
This staged exposure supports gradual abstraction while reinforcing conceptual understanding.

Quantum computational thinking in this tier includes problem decomposition, pattern recognition in quantum algorithms, abstraction through circuit models, and evaluation of computational complexity. 
By the end of Tier~2, students are able to design, implement, and analyse small to medium scale quantum programs in simulation or cloud-based environments.

\subsection{Tier 3: Hardware Aware Development}

Tier~3 introduces the physical realization of quantum computation and the constraints imposed by real hardware~\cite{david2025}. 
Students examine how different qubit technologies, connectivity graphs, coherence properties, and gate sets influence algorithm performance and design choices.

Instruction emphasizes noise aware programming and system level reasoning. 
Learners analyse error sources, perform basic benchmarking, and explore error mitigation and correction techniques appropriate for near term devices. 
Rather than treating hardware as an implementation detail, this tier integrates hardware considerations directly into algorithm design and optimization.

This tier prepares students to reason about trade-offs between algorithm depth, noise tolerance, and resource requirements. 
It also supports learners interested in hardware, architecture, and system integration roles within the quantum ecosystem.

\subsection{Tier 4: Sociotechnical Contextualization}

Tier~4 situates quantum technologies within broader societal, economic, and ethical contexts. 
Students examine the implications of quantum computing for cryptography, cybersecurity, national infrastructure, and emerging industries. 
Topics include responsible innovation, policy considerations, and the limitations of claimed quantum advantage.

This tier also introduces pathways for innovation and impact, including entrepreneurship, intellectual property, and technology transfer. 
Case studies highlight how technical decisions intersect with regulatory frameworks, economic incentives, and societal values~\cite{Srensen2002, Zhang2025}. 
The intent is to develop not only technical expertise, but also informed judgment and professional responsibility.

\subsection{Overlapping Educational Elements}

Three mechanisms operate across all tiers of the QLP. 
Adaptive learning pathways allow students to enter the curriculum at appropriate levels based on background and experience. 
Competency based assessment emphasizes demonstrated understanding over time based progression. 
Authentic project integration ensures that learning outcomes are reinforced through open ended, interdisciplinary work.

Together, these elements enable the QLP to function as a coherent, scalable framework rather than a sequence of disconnected courses. 
The pyramid structure supports continuous reinforcement of core ideas while aligning quantum education with both academic rigour and real world relevance.

\section{Curriculum Architecture}
\label{sec:curriculum}
The Quantum Learning Pyramid (QLP) is realised through a modular, competency based curriculum architecture that supports flexibility, scalability, and alignment with workforce and research needs. 
The curriculum spans introductory undergraduate instruction through advanced graduate specialisation and is designed to accommodate learners with diverse academic backgrounds.

\subsection{Modular Curriculum Design}

The curriculum is organized into a set of core modules (Table~\ref{tab:modules}), each aligned with one or more tiers of the QLP and defined by explicit learning competencies. 
Modules may be offered as stand-alone courses or combined into sequences, enabling institutions to tailor programs to local constraints while preserving curricular coherence.

\begin{table}[h]
\centering
\caption{QLP Core Modules and Associated Competencies}
\begin{tabularx}{0.48\textwidth}{
  >{\raggedright\arraybackslash}p{0.08\textwidth}
  >{\raggedright\arraybackslash}X
  >{\raggedright\arraybackslash}X
  >{\raggedright\arraybackslash}X
}
\toprule
\textbf{Module} & \textbf{Core Concepts} & \textbf{Target Competencies} & \textbf{Credits} \\ \toprule
\textbf{QPhenom} & Superposition, interference, measurement & Design quantum experiments; analyse measurement statistics & 4 \\ \hline
\textbf{QDesign} & Circuits, algorithms, complexity & Design quantum circuits; reason about quantum speed-up & 5 \\ \hline
\textbf{QSoftware} & Programming models, optimization, mitigation & Develop and optimize quantum programs & 6 \\ \hline
\textbf{QHardware} & Qubit technologies, control, characterization & Analyse hardware constraints; design control sequences & 5 \\ \hline
\textbf{QSystems} & Architectures, compilation, error correction & Integrate algorithms with hardware and systems & 6 \\ \hline
\textbf{QSociety} & Ethics, policy, entrepreneurship & Evaluate societal impact; apply responsible innovation & 4 \\ \bottomrule
\end{tabularx}
\label{tab:modules}
\end{table}

Foundational undergraduate courses emphasise conceptual understanding with minimal mathematical prerequisites. 
Early instruction adopts a “Quantum Without Linear Algebra” approach to develop intuition for superposition, entanglement, and measurement. 
As students progress, formal mathematical tools and programming frameworks such as \textit{pennylane}~\cite{pennylane_2022}, or Qiskit are introduced to support algorithm implementation and experimentation.

\subsection{Learning Pathways \& Specializations}

To address diverse learner goals, the curriculum supports multiple specialisation pathways. 
Each pathway draws from the same core modules but emphasises different competencies and depth of coverage:

\begin{itemize}
    \item \textbf{Quantum Theory:} Mathematical foundations, quantum information theory, and advanced algorithms.
    \item \textbf{Quantum Engineering:} Hardware platforms, control systems, noise characterization, and error correction.
    \item \textbf{Quantum Software:} Programming models, compilation, algorithm optimization, and benchmarking.
    \item \textbf{Quantum Applications:} Domain-focused applications such as chemistry, optimization, and machine learning.
\end{itemize}

This structure allows students to share a common conceptual foundation while developing expertise aligned with academic, industrial, or interdisciplinary career paths.

\subsection{Course Progression \& Content Coverage}

Introductory courses introduce quantum principles including wave-particle duality, superposition, entanglement, and probabilistic measurement. 
Students learn qubit operations using single and two qubit gates, explore the circuit model of computation, and study interference through simple algorithmic examples. 
Core algorithms such as Deutsch–Jozsa, Bernstein–Vazirani, Simon’s algorithm, and Grover’s search are used to illustrate quantum advantage. 
A capstone implementation of Shor’s algorithm on a simulator integrates these concepts.

Advanced undergraduate and graduate courses extend this foundation to include quantum error correction, teleportation, cryptography, and NISQ-era (Noisy Intermediate Scale Quantum era), algorithms. 
Graduate offerings typically follow a two course sequence. 
The first builds fluency in programming and algorithm design, while the second focuses on fault tolerance, architectures, and current research challenges. 
Seminar style projects and open ended research problems are incorporated to support deeper specialization.

\subsection{Micro-credentials \& Workforce Alignment}

To support workforce development and lifelong learning, the QLP includes a system of stackable micro-credentials~\cite{Hunt2019}. 
These credentials certify demonstrated competencies and can be earned independently or combined toward formal degree requirements. Examples include Quantum Programming Foundations, Quantum Algorithm Design, Quantum Hardware Operations, and Quantum Applications Specialist.

\subsection{Alignment with ACM/IEEE Guidelines}

The curriculum is explicitly aligned with the ACM/IEEE CS2023 guidelines, including the “AR-Quantum: Quantum Architectures” knowledge unit~\cite{Kumar2024}. 
Learning outcomes map directly to recommended competencies such as implementing quantum algorithms in simulation, analysing hybrid classical-quantum workflows, and reasoning about hardware constraints. 
By aligning course outcomes and assessments with these standards, the QLP ensures that graduates acquire skills consistent with contemporary expectations for quantum ready professionals.

\section{Educational Methodologies}

The Quantum Learning Pyramid (QLP) is implemented through a set of complementary educational methodologies that emphasise active engagement, conceptual grounding, and authentic practice. 
These methods are designed to support the modular curriculum and learner pathways described in Section~\ref{sec:curriculum}. These also address the cognitive and interdisciplinary challenges of quantum education.

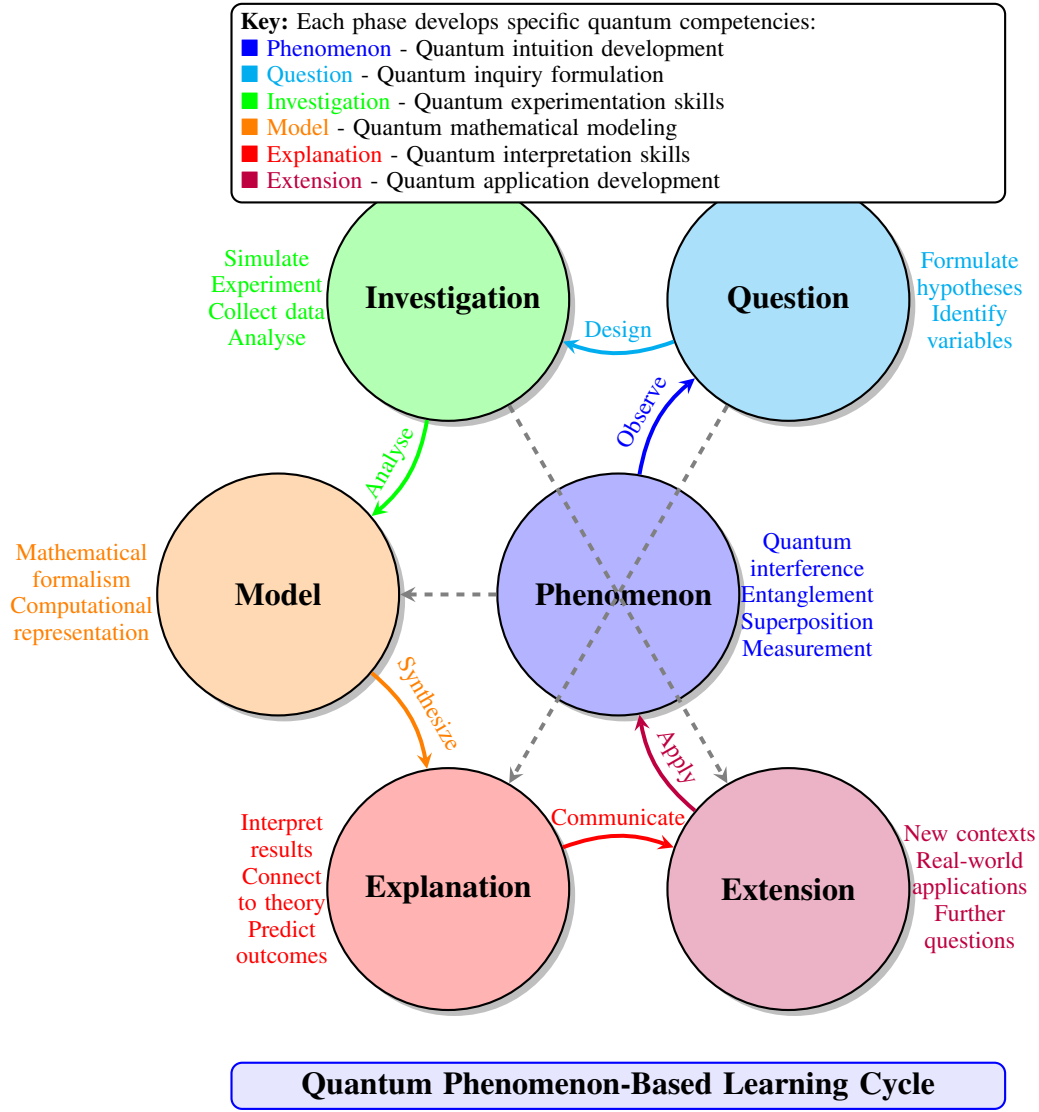
\begin{figure*}[h!]
\centering
\hspace*{-2.3cm}
\begin{tikzpicture}[
    node distance=4cm,
    stage/.style={circle, minimum size=3.2cm, text centered, draw=black, thick, 
                  drop shadow, font=\large\bfseries, text width=2.2cm},
    phase1/.style={stage, fill=blue!30},
    phase2/.style={stage, fill=cyan!30},
    phase3/.style={stage, fill=green!30},
    phase4/.style={stage, fill=orange!30},
    phase5/.style={stage, fill=red!30},
    phase6/.style={stage, fill=purple!30},
    arrow/.style={->, >=stealth, thick, line width=1.5pt},
    desc/.style={text width=2cm, align=center, font=\small}
]

\node[phase1] (phenomenon) at (0:0) {Phenomenon};
\node[phase2] (question) at (60:4.5) {Question};
\node[phase3] (investigation) at (120:4.5) {Investigation};
\node[phase4] (model) at (180:4.5) {Model};
\node[phase5] (explanation) at (240:4.5) {Explanation};
\node[phase6] (extension) at (300:4.5) {Extension};

\draw[arrow, blue] (phenomenon) to [bend left=20] node[midway, above, sloped, font=\small] {Observe} (question);
\draw[arrow, cyan] (question) to [bend left=20] node[midway, above, sloped, font=\small] {Design} (investigation);
\draw[arrow, green] (investigation) to [bend left=20] node[midway, above, sloped, font=\small] {Analyse} (model);
\draw[arrow, orange] (model) to [bend left=20] node[midway, above, sloped, font=\small] {Synthesize} (explanation);
\draw[arrow, red] (explanation) to [bend left=20] node[midway, above, sloped, font=\small] {Communicate} (extension);
\draw[arrow, purple] (extension) to [bend left=20] node[midway, above, sloped, font=\small] {Apply} (phenomenon);

\node[desc, blue] at ([xshift=2.5cm, yshift=0.0cm] phenomenon) {Quantum interference\\Entanglement\\Superposition\\Measurement};
\node[desc, cyan] at ([xshift=2.4cm, yshift=0.0cm] question) {Formulate\\hypotheses\\Identify\\variables};
\node[desc, green] at ([xshift=-2.4cm, yshift=0.0cm] investigation) {Simulate\\Experiment\\Collect data\\Analyse};
\node[desc, orange] at ([xshift=-2.6cm, yshift=0.0cm] model) {Mathematical\\formalism\\Computational\\representation};
\node[desc, red] at ([xshift=-2.2cm, yshift=0.0cm] explanation) {Interpret results\\Connect to theory\\Predict outcomes};
\node[desc, purple] at ([xshift=2.4cm, yshift=0.0cm] extension) {New contexts\\Real-world\\applications\\Further questions};

\draw[arrow, dashed, gray] (phenomenon) -- (model);
\draw[arrow, dashed, gray] (question) -- (explanation);
\draw[arrow, dashed, gray] (investigation) -- (extension);

\node[rectangle, fill=blue!10, draw=blue, thick, rounded corners, 
      text width=10cm, align=center, font=\large\bfseries] 
      at (0, -6.5) {Quantum Phenomenon-Based Learning Cycle};

\node[rectangle, fill=white, draw=black, thick, rounded corners, 
      text width=10cm, align=left, font=\small] 
      at (0, 6.5) {
      \textbf{Key:} Each phase develops specific quantum competencies: \\
      \textcolor{blue}{$\blacksquare$ Phenomenon} - Quantum intuition development \\
      \textcolor{cyan}{$\blacksquare$ Question} - Quantum inquiry formulation \\
      \textcolor{green}{$\blacksquare$ Investigation} - Quantum experimentation skills \\
      \textcolor{orange}{$\blacksquare$ Model} - Quantum mathematical modeling \\
      \textcolor{red}{$\blacksquare$ Explanation} - Quantum interpretation skills \\
      \textcolor{purple}{$\blacksquare$ Extension} - Quantum application development
      };

\end{tikzpicture}
\caption{Phenomenon-based learning cycle for quantum education.}
\label{fig:phenomenon}
\end{figure*}

\subsection{Phenomenon-Based Learning}

Instruction across all tiers of the QLP is anchored in quantum phenomena rather than abstract formalism. 
Observable effects such as interference, entanglement, and measurement outcomes serve as entry points for inquiry and reasoning. 
Learning activities are organized around a structured inquiry cycle, illustrated in Fig.~\ref{fig:phenomenon}, which progresses through observation, questioning, investigation, modelling, explanation, and extension.
\[
\begin{aligned}
\mathcal{C}_{\text{inquiry}} =
\{\text{Phenomenon} \rightarrow \text{Question}
\rightarrow \text{Investigation} \\ \rightarrow \text{Model}
\rightarrow \text{Explanation} \rightarrow \text{Extension}\}
\end{aligned}
\]

This cycle supports gradual abstraction by allowing students to first build intuition, then formalize their understanding using mathematical and computational models. 
The extension phase explicitly connects core concepts to new contexts and applications, reinforcing transfer of learning.

\subsection{Hybrid Reality Laboratories}

Laboratory experiences are delivered through a hybrid reality model that integrates physical experimentation, high-fidelity simulation, and visualization tools. 
Where feasible, students interact with tabletop quantum setups or remotely accessible hardware. 
Virtual laboratories provide scalable access to quantum systems through simulators and cloud based platforms. 
Augmented and interactive visualizations are used to represent quantum states, circuits, and measurement statistics that are otherwise inaccessible.

This blended approach ensures broad accessibility while preserving experimental authenticity. 
It also allows laboratories to scale in complexity as students move upward through the QLP tiers.

\subsection{Adaptive Learning \& Personalization}

To accommodate diverse backgrounds and learning rates, the QLP incorporates adaptive learning mechanisms that personalise instructional content and practice opportunities. 
Learning analytics are used to identify conceptual gaps and recommend targeted activities. 
Conceptually, the adaptation strategy balances mastery and exploration:
\[
A_t(s) = \arg\max_a \left[ Q(s,a) + U(s,a) \right]
\]
where $Q(s,a)$ represents expected learning gain and $U(s,a)$ encourages exposure to underdeveloped concepts. This approach supports competency based progression without enforcing uniform pacing.

\subsection{Project Based Quantum Challenges}

Project based learning (PBL) is a central component of the QLP. 
Students engage in open ended, interdisciplinary challenges drawn from current research and industry practice. 
Example projects include quantum chemistry simulations for materials discovery~\cite{Bauer2020}, 
portfolio optimisation using quantum algorithms~\cite{Kerenidis2019}, quantum machine learning for medical diagnostics~\cite{Ullah2024}, 
and secure communication protocols using quantum networks~\cite{Hildebrand2025}.

Projects require students to formulate problems, implement algorithms, analyse results, and communicate findings. 
Early projects emphasise guided exploration, and advanced projects focus on research or capstone experiences. 
This progression strengthens conceptual retention and professional readiness.

\subsection{Integrated Active Learning Practices}

Weekly laboratory exercises complement lectures and projects. 
Early labs focus on circuit construction and measurement analysis using tools such as Qiskit or Cirq, helping students confront misconceptions through direct experimentation. 
As students advance, labs evolve into full algorithm implementation, debugging, and performance evaluation.

Formal technical communication is embedded throughout the curriculum. 
Students produce short conference style reports and participate in structured peer review. 
Many projects involve dual mentorship across disciplines, reflecting the collaborative nature of quantum research and development. 
These practices strengthen communication skills and reinforce research norms.

Additional active learning techniques, including in-class problem solving, flipped instruction, and interactive simulations, are used to support non-intuitive concepts. 
Visual representations such as Bloch spheres, probability distributions, and circuit animations help connect mathematical expressions to observable behaviour. 
Together, these methodologies ensure that learning within the QLP is experiential, rigorous, and aligned with real world quantum practice.

\section{Assessment Framework}
\label{sec:assessment}
Assessment within the QLP is designed to measure conceptual understanding, practical skill development, and the ability to apply quantum knowledge in authentic contexts. 
Consistent with the competency based curriculum architecture, an assessment that emphasises demonstrated mastery rather than time based completion is proposed. 
Multiple assessment modes are used to capture different dimensions of learning and to support continuous feedback.

\subsection{Multidimensional Assessment Design}
The QLP adopts a multidimensional assessment strategy that integrates formative, summative, and authentic evaluations across conceptual, computational, experimental, and sociotechnical domains (Table~\ref{tab:matrix}). 
This structure ensures alignment between learning objectives, instructional activities, and assessment outcomes.

\begin{table}[h!]
\centering
\caption{QLP Multidimensional Assessment Matrix}
\label{tab:matrix}
\begin{tabularx}{0.45\textwidth}{
  >{\raggedright\arraybackslash}p{0.09\textwidth}
  >{\raggedright\arraybackslash}X
  >{\raggedright\arraybackslash}X
  >{\raggedright\arraybackslash}X
}
\toprule
\textbf{Learning Dimension} & \textbf{Formative Assessment} & \textbf{Summative Assessment} & \textbf{Authentic Assessment} \\ \toprule
\textbf{Conceptual} & Concept maps, peer instruction & Validated concept inventories (e.g., QCCS, QISCIT) & Explain and predict quantum phenomena \\ \hline
\textbf{Computational} & Circuit design exercises & Algorithm implementation & Optimize and analyse quantum circuits \\ \hline
\textbf{Experimental} & Lab notebooks, data analysis & Experimental design tasks & Characterize quantum systems \\ \hline
\textbf{Sociotechnical} & Ethical reflections, policy briefs & Impact analyses & Develop responsible use guidelines \\ \bottomrule
\end{tabularx}
\end{table}

Formative assessments provide rapid feedback and support iterative improvement, while summative assessments evaluate overall competency at key transition points. 
Authentic assessments mirror professional quantum practice and require students to integrate knowledge across multiple domains.

\subsection{Conceptual Assessment Using Validated Inventories}

The QLP assesses conceptual understanding of core threshold concepts, including superposition, entanglement, measurement, and the quantum-classical interface. Rather than introducing a new assessment instrument, the framework leverages existing validated quantum concept inventories appropriate to the course level. Suitable examples include the Quantum Computing Conceptual Survey (QCCS) for introductory quantum computing courses~\cite{PhysPortQCCS,Meyer2022} and the Quantum Information Science Concept Inventory Test (QISCIT), which is designed for learners without extensive mathematical prerequisites~\cite{Durkin2025}.

When conceptual scores from multiple subscales or inventories are combined, a weighted composite score may be used:
\[
S=\sum_{i=1}^{n} w_i C_i \ ,
\]
where $C_i$ denotes the competency measured by a validated instrument and $w_i$ represents its relative instructional importance. This composite provides a simple mechanism for diagnosing misconceptions and monitoring conceptual development across the QLP curriculum while relying on established, peer-reviewed assessment instruments.

\subsection{Qualitative Evaluation and Feedback}

Quantitative measures are complemented by qualitative evaluation. 
Thematic analysis of student reflections, peer feedback, and instructor observations are expected to indicate increased engagement
under the \emph{phenomenon first} approach and improved \emph{conceptual coherence} through spiral progression. 
Students tend to report higher self efficacy when assessed on demonstrated competencies rather than isolated exams~\cite{Vrugt1997, vanDinther2014}. 
Instructors may note coordination challenges in interdisciplinary teaching, highlighting the need for structured collaboration and shared assessment rubrics.

Together, these assessment practices propose a comprehensive view of student learning and ensure alignment between the QLP framework, curriculum architecture, and educational methodologies.
\section{Discussion}
\label{sec:discussion}
This section discusses the anticipated outcomes of the Quantum Learning Pyramid (QLP) framework and positions it relative to existing quantum education efforts and to the empirical evaluation it still requires.
At this stage, the QLP is a conceptual proposal: no implementation data yet exist, and the discussion below is framed accordingly, as anticipated outcomes and a proposed evaluation plan rather than as measured results.

Existing evidence on present quantum computing curricula is limited but informative. Surveys of quantum information science instructors report highly heterogeneous course content, prerequisites, and assessment practice across institutions~\cite{Meyer2022}, and studies of student learning in introductory quantum computing courses document persistent, specific misconceptions about superposition, measurement, and entanglement that survive standard instruction~\cite{hu2024}. This heterogeneity and these persistent misconceptions motivate the QLP's emphasis on phenomenon-first instruction, spiral progression, and a shared conceptual assessment, as described in the preceding sections, but they do not by themselves demonstrate that the QLP improves on current practice.
\subsection{Pedagogical Contributions}
Assessment proposes several instructional contributions of the QLP framework. 
First, the simultaneous vertical and horizontal progression of content, implemented through a spiral structure, supports both conceptual deepening and application breadth. 
Students can improve coherence across foundational and advanced topics, as reflected in longitudinal gains on validated concept-inventory scores and performance on integrative assessments.

Second, explicit scaffolding around quantum threshold concepts is essential in reducing persistent misconceptions, and QLP provides that.
Targeted formative assessments, such as concept mapping and peer instruction, could enable early identification of conceptual barriers and informed instructional adjustments.

Third, the hardware--software co-design pedagogy is expected to strengthen student's ability to reason across abstraction layers. 
Authentic assessments requiring circuit optimisation and system characterisation will increase fluency in connecting physical constraints with algorithmic design choices.

Finally, embedding sociotechnical considerations throughout the curriculum, rather than isolating them in stand-alone modules, leads to more nuanced ethical reasoning. 
Students are expected to demonstrate the ability to assess societal impact and articulate in capstone style assessments, indicating meaningful integration of technical and societal competencies.
\subsection{Positioning Relative to Existing Quantum Education Initiatives}
The QLP's contribution lies primarily in integrating established educational approaches into a single tiered, competency based architecture for quantum information science and computing learning curriculum. A substantial and growing set of programs already address parts of this space. Vendor and community platforms, including hardware-vendor training materials, QWorld, and related workshop and hackathon based initiatives, provide accessible introductory training and outreach at scale~\cite{Metwalli2026, Kaur2022}. Multilateral efforts such as the Open Quantum Institute, pair cloud hardware access with capacity-building activities aimed at underserved regions in support of the UN Sustainable Development Goals~\cite{OQI2026}. At the degree-program level, dedicated undergraduate quantum engineering curricula have recently been established, for example through a multi-institution collaboration~\cite{Asfaw2022} and at the University of New South Wales~\cite{Dzurak2022}; both combine physics, hardware, and systems content in a manner similar in spirit to Tiers~1-3 of the QLP.

Building on these efforts, the QLP provides an instructional framework that integrates multiple pedagogical approaches tailored to the unique requirements of QIS\&C. Active learning, project based learning, and phenomenon-first instruction are all individually well established, and several are already used within the programs above. The distinguishing feature of the QLP is the integration of established instructional practices into a coherent four-tier learning progression supported by the competency-based, multidimensional assessment framework described in Section~\ref{sec:assessment}. The framework simultaneously addresses conceptual, computational, experimental, and sociotechnical competencies while remaining adaptable to undergraduate, graduate, workforce development, and outreach settings. Existing community platforms, vendor training programs, and academic curricula primarily focus on content delivery, hands-on experience, or institution-specific course sequencing. In contrast, the QLP unifies curriculum progression, assessment, and learner adaptation within a single framework. This contribution lies in the integration of these elements, and its effectiveness requires future empirical evaluation.
\subsection{Scalability \& Contextual Adaptation}
The modular and competency based nature of the QLP enables adaptation across educational contexts without compromising assessment integrity. 
Simplified conceptual modules and qualitative assessments support secondary level instruction, and concentrated, role specific modules are suitable for professional development and workforce reskilling. 
For public outreach and informal education, phenomenological demonstrations paired with reflective assessments allow engagement without extensive mathematical prerequisites.

Across these settings, the assessment framework provides a common structure for measuring learning outcomes, enabling consistent evaluation while allowing curricular customisation.
\subsection{Industry Readiness \& Professional Skill Development}
Project based assessments mirror professional quantum research and development workflows, requiring students to define requirements, document designs, justify trade-offs, and communicate results. 
These competencies extend beyond technical proficiency to include collaboration, technical writing, and interdisciplinary reasoning.
This indicate strong alignment between curricular outcomes and industry expectations. 

Assessments, such as applied optimization problems or quantum machine learning case studies, expose students to real world constraints and current hardware platforms. 
Guest lectures, internships, and collaborations further reinforce industry relevance and provide external validation of learning outcomes.

To maintain long term alignment, the curriculum and assessment instruments are to be periodically reviewed against ACM/IEEE guidelines and feedback from industry partners and program alumni. 
This iterative process ensures continued relevance as quantum technologies and workforce demands evolve.
\subsection{Minimum Viable Implementation}
Although the QLP is presented as a comprehensive framework, it is intended to support incremental adoption. A minimum viable implementation (MVI) can be deployed within a single existing course by one instructor, focusing on Tier~1 and Tier~2 (Section~\ref{sec:qlp_frmwrk}). Such an implementation requires only freely available quantum software platforms (e.g., pennylane, Qiskit or Cirq), one project-based assignment, and a validated quantum concept inventory administered before and after instruction. Learning gains may be evaluated through pre/post assessments, student reflections, or comparison with a parallel offering when available.

The QLP can therefore be viewed as a collection of reusable educational components rather than a curriculum requiring complete adoption. Its tiered learning model, modular curriculum, and competency-based assessment framework can be adopted independently and adapted to local institutional needs. This modular approach lowers the barrier to implementation while encouraging the community to share instructional materials, assessment instruments, laboratory activities, and implementation experiences to support the continued evolution of quantum computing education.

\subsection{Limitations \& Future Directions}

As a primarily conceptual proposal, the framework requires systematic experimental evaluation before its effectiveness can be fully assessed. Despite this, certain limitations can be outlined at this stage.
Effective implementation requires instructor expertise spanning multiple fields including physics, computer science, and engineering, demanding the need for targeted faculty development. 
While cloud based platforms mitigate some costs, access to experimental hardware remains a barrier for many institutions. 
Evaluating long term career impact could be challenging, motivating the need for longitudinal studies linking assessment outcomes to professional trajectories.

Future work will focus on the minimum viable implementation described above, together with the following:
\begin{itemize}
    \item Piloting the MVI with volunteer instructors, using an existing validated concept inventory (QCCS or QISCIT) as a pre/post outcome measure
    \item Expanding international collaboration to harmonise quantum education standards
    \item Developing and sharing open quantum education resources, rubrics, and laboratory configurations as collective infrastructure
    \item Establishing a quantum education research network to support evidence based pedagogy
    \item Strengthening industry-academia partnerships, enabling co-designed curricula and assessment practices, that ensure quantum education evolves in step with technological and societal change
\end{itemize}

\section{Conclusion}

This paper presented the Quantum Learning Pyramid (QLP) as a comprehensive, research informed framework for quantum information and computing education. 
The QLP integrates phenomenological learning, computational thinking, hardware awareness, and sociotechnical context into a coherent instructional model. 
By doing so, it directly addresses persistent challenges in quantum education, including abstraction barriers, fragmented curricula, and gaps between theory and practice.

The proposed competency based, modular curriculum spans undergraduate and graduate levels and aligns with current ACM/IEEE curriculum guidelines. 
Active pedagogies, including laboratory exercises, project based learning, and authentic assessment, are embedded throughout. 
Modern tools such as quantum simulators and cloud accessible quantum processors support hands-on learning and reinforce conceptual understanding. 
Together, these elements provide a structured pathway for developing quantum competencies from first principles to advanced applications.

While empirical validation remains a future work, the QLP demonstrates a practical and adaptable approach to quantum education. 
The framework supports diverse learner pathways and enables alignment with both academic standards and industry expectations. 
As quantum technologies continue to transition from research to deployment, educating a skilled workforce is critical.
Educational frameworks such as the QLP will be essential for preparing a quantum ready workforce and fostering a scientifically literate society that is capable of engaging with the opportunities and implications of the quantum era.
\vspace{-0.1in}

\bibliographystyle{IEEEtran}
\bibliography{ref.bib}

\end{document}